 \documentclass[preprint,2012,12pt]{elsarticle}

 \usepackage{graphics}
 \usepackage{longtable}
 \usepackage{graphicx}
 \usepackage{epsfig}

\usepackage{amssymb}
 \usepackage{amsthm}
 \usepackage{booktabs}

\journal{Radiat. Phys. Chem.}

\begin{document}

\hyphenation{ener-gy ave-ra-ge Ave-ra-ge ge-ne-ra-ted re-la-ti-vi-stic dif-fe-ren-ce in-sensi-ti-ve ki-ne-ma-ti-cal Sy-ste-ma-tic sy-ste-ma-tic ave-ra-ging con-ver-ters con-ver-ter la-bo-ra-to-ry se-con-da-ry
se-con-da-ries si-mu-la-tion si-mu-la-tions do-mi-na-te do-mi-na-tes ta-bu-la-tions
Ta-bu-la-tions Di-stri-bu-tions
di-stri-bu-tions di-stri-bu-tion Di-stri-bu-tion Di-spla-ce-ment di-spla-ce-ment Di-spla-ce-ments Bethe power powers ma-te-rial ma-te-rials
di-spla-ce-ments Li-near li-near Ca-sca-de Ca-sca-des ca-sca-de ca-sca-des
ta-bu-la-tion Ta-bu-la-tion re-pe-ti-ti-ve E-lec-tron E-lec-trons
e-lec-tron e-lec-trons De-tec-tion Pro-duc-tion pro-duc-tion
Re-so-lu-tions Re-so-lu-tion re-so-lu-tions re-so-lu-tion
Ope-ra-tion mi-ni-mum Ener-gy Ener-gies fer-ro-mag-net
fer-ro-mag-nets meta-sta-ble meta-sta-bi-lity con-fi-gu-ra-tion
con-fi-gu-ra-tions expo-nen-tially mo-bi-li-ty- mo-bi-li-ties
tem-pe-ra-tu-re tem-pe-ra-tu-res con-cen-tra-tion con-cen-tra-tions
elec-tro-nic elec-tro-nics STMelec-tro-nics sec-tion Sec-tion
Chap-ter chap-ter theo-ry ap-pro-xi-mation ra-dia-tion Ra-dia-tion
ca-pa-ci-tan-ce approaches tran-sport dispersion Ca-lo-ri-me-try
ca-lo-ri-me-try En-vi-ron-ment En-vi-ron-ments en-vi-ron-ment
en-vi-ron-ments Fur-ther-mo-re do-mi-nant ioni-zing pa-ra-me-ter pa-ra-me-ters
O-sa-ka ge-ne-ral exam-ple Exam-ple ca-vi-ty Ca-vi-ty He-lio-sphe-re
he-lio-sphe-re dis-tan-ce Inter-pla-ne-ta-ry inter-pla-ne-ta-ry
ge-ne-ra-li-zed sol-ving pho-to-sphe-re sym-me-tric du-ring
he-lio-gra-phic strea-ming me-cha-nism me-cha-nisms expe-ri-mental
Expe-ri-mental im-me-dia-tely ro-ta-ting na-tu-rally
ir-re-gu-la-ri-ties o-ri-gi-nal con-si-de-red e-li-mi-na-ting
ne-gli-gi-ble stu-died dif-fe-ren-tial mo-du-la-tion ex-pe-ri-ments
ex-pe-ri-ment Ex-pe-ri-ment Phy-si-cal phy-si-cal in-ve-sti-ga-ted
Ano-de Ano-des ano-de ano-des re-fe-ren-ce re-fe-ren-ces
ap-pro-xi-ma-ted ap-pro-xi-ma-te in-co-ming bio-lo-gi-cal
atte-nua-tion other others eva-lua-ted nu-cleon nu-cleons reac-tion
pseu-do-ra-pi-di-ty pseu-do-ra-pi-di-ties esti-ma-ted va-lue va-lues
ac-ti-vi-ty ac-ti-vi-ties bet-ween Bet-ween dis-cre-pan-cy
dis-cre-pan-cies cha-rac-te-ri-stic
cha-rac-te-ri-stics sphe-ri-cally anti-sym-metric ener-gy ener-gies
ri-gi-di-ty ri-gi-di-ties leaving pre-do-mi-nantly dif-fe-rent
po-pu-la-ting acce-le-ra-ted respec-ti-ve-ly sur-roun-ding
sa-tu-ra-tion vol-tage vol-tages da-ma-ge da-ma-ges be-ha-vior
equi-va-lent si-li-con exhi-bit exhi-bits con-duc-ti-vi-ty
con-duc-ti-vi-ties dy-no-de dy-no-des created Fi-gu-re Fi-gu-res
tran-si-stor tran-si-stors Tran-si-stor Tran-si-stors ioni-za-tion
Ioni-za-tion ini-tia-ted sup-pres-sing in-clu-ding maxi-mum mi-ni-mum
vo-lu-me vo-lu-mes tu-ning ple-xi-glas using de-pen-ding re-si-dual har-de-ning li-quid
know-ledge usage me-di-cal par-ti-cu-lar scat-te-ring ca-me-ra se-cond hea-vier hea-vy trans-axial
con-si-de-ration created Hy-po-the-sis hy-po-the-sis usually inte-ra-ction Inte-ra-ction
inte-ra-ctions Inte-ra-ctions pro-ba-bi-li-ty pro-ba-bi-li-ties
fol-low-ing cor-re-spon-ding e-la-stic readers reader pe-riod pe-riods geo-mag-ne-tic sa-ti-sfac-tory ori-gi-nal-ly
fac-to-ri-za-tion}

\begin{frontmatter}



\title{AN EXPRESSION FOR THE MOTT CROSS SECTION OF ELECTRONS AND POSITRONS ON NUCLEI WITH $Z$ UP TO 118\\ \small{(Accepted for publication in Radiation Physics and Chemistry)}}


\author[1,2]{M.J Boschini}
\author[1]{C. Consolandi}
\author[1,3]{M. Gervasi}
\author[4]{S. Giani}
\author[1]{D. Grandi}
\author[4]{V. Ivanchenko}
\author[5]{P. Nieminem}
\author[1,3]{S. Pensotti}
\author[1]{P.G. Rancoita}
\author[1,3]{M. Tacconi}

\address[1]{Istituto Nazionale di Fisica Nucleare, INFN Milano-Bicocca, Milano (Italy)}
\address[2]{CINECA, Segrate (MI) (Italy)}
\address[3]{Department of Physics, University of Milano Bicocca, Milano (Italy)}
\address[4]{CERN, Geneva, 23, CH-1211, Switzerland}
\address[5]{ESA, ESTEC, AG Noordwijk (Netherlands)}

\begin{abstract}
In the present work, an improved numerical solution for determining the
ratio,$\mathcal{R}^{\rm Mott}$, of the unscreened Mott differential cross section (MDCS) with respect to Rutherford's formula is proposed for the scattering of electrons and positrons on nuclei with $1\leq Z \leq 118$.~It accounts for incoming lepton energies between 1\,keV and 900\,MeV.~For both electrons and positrons, a fitting formula and a set of fitting coefficients for the ratio $\mathcal{R}^{\rm Mott}$ on nuclei are also presented.~The found average error of the latter practical interpolated expression is typically lower than 1\% also at low energy for electrons and lower than 0.05\% for positrons for all nuclei over the entire energy range.
\par
Both the improved numerical solution and the interpolated practical expression were found in good agreement with the partially available previous calculations.
\end{abstract}


\end{frontmatter}


\section{Introduction}
Mott~(1929; 1932) (in addition, see the discussion available in Sections~4--4.5 of Chapter~IX of~\cite{Mott_book}) treated the scattering of electrons by unscreened and infinitively heavy nuclei with negligible spin effects, by extending a method of Wentzel~(1927) (see also~\cite{Born1}) and including effects related to the spin of electrons.~Wentzel's method was dealing with incident and scattered waves on point-like nuclei.~The differential cross section (DCS) -~the so-called \textit{Mott} (\textit{unscreened}) \textit{differential cross section} (MDCS) - was expressed by Mott~(1929; 1932) as two conditionally convergent infinite series in terms of Legendre expansions (see also Bartlett and Watson~(1940) and Equation 46 in Chapter~IX of~\cite{Mott_book}).~In Mott--Wentzel treatment, the scattering occurs on a field of force generating a radially dependent Coulomb - unscreened [screened] Mott~(1929; 1932) [Wentzel~(1927)] - potential.~Mott equation - computed using Darwin's solution to the Dirac equation - is also referred to as an \textit{exact} formula for the differential cross section, because no Born approximation of any order is used in its determination.
\par
It has to be remarked that Mott's treatment of collisions of fast electrons with atoms - accounting for screening effects - involves the knowledge of the wave function of the atom and uses the first Born approximation (e.g.,~see Sections 2--5 of Chapter~XVI of~\cite{Mott_book}), thus, as discussed by many authors (for instance, see \cite{Idoeta,Lijian,Consolandi_2012}, see also references therein), in most cases the computation of the cross section depends on the application of numerical methods.~Particularly in calculations for electron transport in materials or in the determination of induced radiation damage due to atomic displacements resulting from Coulomb interaction on nuclei, this treatment may require an excessive time-consuming procedure for accounting the effect of nuclear screening by atomic electrons.~
\par
In practice for the above mentioned calculations (e.g.,~see \cite{Cahn,Seitz,Fernandez1,Butkevick,geant4,Sempau,rop_si,Jun_2009,Consolandi_2012} and references therein, see Chapter~4 of \cite{LR_3rd}), a factorization of the elastic screened cross section is often employed (e.g.,~see \cite{Zeitler,Berger_comp,Idoeta,Fernandez,Lijian,Consolandi_2012}, Chapter~1 of \cite{SSRD} and references therein).~It involves the unscreened differential cross section on point-like nuclei and a factor which takes into account the screening of the nuclear charge by the atomic electrons.~Expressions for this term - which is also employed in the treatment of nucleus--nucleus interactions - were derived and discussed by many authors (e.g., see \cite{Wentzel,Moliere,Spencer,m_sca_other,Sherman,Zeitler,Salvat,Idoeta,Fernandez1,Fernandez,Butkevick,Boschini,Boschini_2011,Consolandi_2012} and references therein, see also  Chapters~2 and~7 of~\cite {Berger_book}, Chapter~2 of~\cite{LR_3rd} and Chapter~1 of~\cite{SSRD}).~Furthermore, in electron scattering on nuclei above 10\,MeV, as discussed by Fernandez-Varea, Mayol and Salvat 1993a (see also \cite{Consolandi_2012} and references therein), the effects due to the finite nuclear size have to be taken into account and are usually expressed by a multiplicative term, the so-called \textit{nuclear form factor} (e.g., see \cite{Helm,Ho57,Nagarajan,DeVries,Bertulani,Duda,Jentschura} and references therein).
\par
Approximated expressions for the Mott (unscreened) differential cross section were derived as early as in the 1940s and 1950s (e.g.,~see \cite{Bartlett_Watson,McKinley,Feshbach1,Curr,Doddett,Sherman}).~Recently, Idoeta and Legarda~(1992) (as suggested, for instance, in~\cite{Sherman}) evaluated the MDCS exploiting recursion relationships of the gamma functions showing that the ratio - appearing in the MDCS - fulfills the condition for the application of the Stirling's formula. In addition, they applied the trasformation of Yennie, Ravenhall and Wilson (1954) to the infinite series of Legendre polynomials.~Finally, they obtained tabulated values for electrons and positrons scattering on a few nuclei with energies from 5\,keV and 10\,MeV and a maximum error of less than $10^{-3}$\,\%.~Subsequently, Lijian, Quing and Zhengming~(1995) developed a fitting procedure for the numerical values determined following the approach of Idoeta and Legarda~(1992), then expressing the ratio ($\mathcal{R}^{\rm Mott}$) of the MDCS to Rutherford differential cross section (RDCS) as an analytical formula depending on 30 parameters with a maximum error of less than 1\,\% only for electrons with energies from 1\,keV up to 900\,MeV.~Above 900\,MeV, no further energy dependence was exhibited by the parameters.~These parameters depend on the nuclear target and were tabulated for nuclei with $Z$ up to 90.
\par
In the current article, the results obtained adapting both the approach, in Sect.~\ref{INAUMDCS}, of Idoeta and Legarda~(1992) (and also ~\cite{Sherman}) and, in Sect.~\ref{PIEUMDCS}, the procedure of Lijian, Quing and Zhengming~(1995) are reported for both electrons and positrons with energies from 1\,keV up to 900\,MeV scattered by nuclei with $Z$ up to 118.~The results from the current improved numerical approach (Sect.~\ref{INAUMDCS}) are compared with what determined by Idoeta and Legarda~(1992) who, in turn, already discussed the good agreement of what they obtained with those previously found - within the sensitivity of the used approximations - in Refs.~\cite{McKinley,Curr,Yadav,Motz_rmp,Doddett,Sherman}.~Those regarding the present practical interpolated expression for the unscreened MDCS (Sect.~\ref{PIEUMDCS}) are compared with what found in Refs.~\cite{Curr,Idoeta,Lijian}.
\par
Finally, it has to be remarked that the described treatment of the MDCS for elastic scattering on nuclei is implemented into Geant4 distribution (e.g., see Ref.~\cite{geant4}) version 9.6 (see, also, Ref.~\cite{Consolandi_2012} and references rherein).

\section{The Unscreened Mott Cross Section}
\label{MDCR_Sect}
As already mentioned, the scattering of electrons (or positrons) on unscreened atomic nuclei with charge number $Z$ was obtained by Mott~(1929; 1932) (see also Sections~4--4.5 of Chapter~IX of~\cite{Mott_book}, Section 1.3.1 of~\cite{SSRD} and references therein), who derived the differential cross section (${d\sigma^{\rm Mott}}/{d\Omega}$ with $d\Omega$ the unit of solid angle) - usually termed as \textit{Mott differential cross section} (MDCS) -  following a treatment in which effects related to the spin of the incoming electron or positron were included.~The MDCS was obtained in the laboratory system of reference for infinitely heavy nuclei initially at rest neglecting effects due to their spin.~In addition, effects related to the recoil and finite rest mass ($M$) of the target nucleus were disregarded and, as a consequence, in this framework the total energy of electrons (or positrons) has to be smaller\footnote{For an ultra high-energy extension of the treatment of the differential cross sections on nuclei, one may see the discussion in~\cite{Consolandi_2012}, Chapter~1 of~\cite{LR_3rd}, Section 1.3.1 of~\cite{SSRD} and references therein. } than $Mc^2$ (where c is the speed of light).
\par
The MDCS, or an approximated expression of it, is commonly\footnote{One can see, for instance, the discussion in~\cite{Feshbach1,Curr,Doddett,Sherman,Idoeta,Lijian,Consolandi_2012} (see, also, references therein).} formulated in terms of its ratio, $\mathcal{R}^{\rm McF}$, with respect to that for a Rutherford scattering,~i.e., the \textit{Rutherford differential cross section} (RDCS) - also termed \textit{Rutherford's formula} (${d\sigma^{\rm Rut}}/{d\Omega}$) (e.g.,~see a treatment in Section 1.6.1 of~\cite{LR_3rd}) - given by:
\begin{eqnarray}
  \label{Rutherd_c_s_com_plab}  \frac{d\sigma^{\rm Rut}}{d\Omega}  & = &
 \left(\! \frac{zZe^2}{ p \beta c}\!\right)^2  \frac{1}{\left(1-\cos \theta \right)^2}\\
  \label{Rutherd_c_s_com_plab_n} &  =&   \left(z Z r_e \right)^2 \left(\!\frac{1-\beta^2}{\beta^4} \!\right) \frac{1}{\left(1-\cos \theta \right)^2},
\end{eqnarray}
where $p$ and $\theta$ are the momentum and scattering angle of the electron (or positron), respectively; $\beta = v/c$ with $v$ the electron (positron) velocity; $z=-1$ ($z=+1$) is the charge number of the electron (positron); finally, $r_e = e^2/(mc^2)$ is the classical electron (or positron) radius with $m$ and $e$ the rest mass and charge of electron (or positron), respectively.~The MDCS is usually expressed in terms of Rutherford's formula as:
\begin{equation}\label{MDCS_Mott_R}
   \frac{d\sigma^{\rm Mott}}{d\Omega}  = \frac{d\sigma^{\rm Rut}}{d\Omega} \,\, \mathcal{R}^{\rm Mott},
\end{equation}
where $\mathcal{R}^{\rm Mott}$ (as above mentioned) is the ratio between the MDCS and RDCS.~
\par
$\mathcal{R}^{\rm Mott}$ can be formulated (e.g.,~see Equation~(2) in~\cite{Sherman}, Equation~(2) in~\cite{Idoeta} and Equation~(1) in~\cite{Lijian}) formulated in terms of the two conditionally convergent series - $F$ and $G$, defined as an expansion in Legendre polynomials - derived by Mott~(1929; 1932) for expressing the MDCS,~i.e.,
\begin{equation}
  \label{R_Mott_Idoeta} \mathcal{R}^{\rm Mott}=\frac{2(1-\cos\theta)}{(\tau+1)^2}|F|^2+\frac{|G|^2}{(\alpha zZ)^2}\frac{2p^2(1-\cos\theta)^2}{(\tau+1)^2(1+\cos\theta)}
\end{equation}
with $\tau$ the kinetic energy expressed of the incoming lepton in units of its rest mass ($m$) (thus, $\tau = \gamma -1$ where $\gamma $ is the \textit{Lorentz factor}) and $\alpha$ the \textit{fine structure constant}.~The complex functions $F$ and $G$ are given by
\begin{eqnarray}
 \label{F_function} F&=&F_0+F_1\\
 \label{G_function} G&=&G_0+G_1
\end{eqnarray}
with
\begin{eqnarray}
 \label{F0_function} F_0&=&-\frac{R}{\mathrm{i}q},\\
 \label{G0_function} G_0&=&R\,\cot^2\frac{\theta}{2},\\
 \label{F1_function} F_1&=&\frac{\mathrm{i}}{2}\sum_{k=0}^{+\infty}[kD_k+(k+1)D_{k+1}](-)^k\mathrm{P}_k(\cos\theta),\\
 \label{G1_function} G_1&=&\frac{\mathrm{i}}{2}\sum_{k=0}^{+\infty}[k^2D_k-(k+1)^2D_{k+1}](-)^k\mathrm{P}_k(\cos\theta),
\end{eqnarray}
where $\mathrm{P}_k(\cos\theta)$ is the Legendre polynomial of order $k$ and
\begin{eqnarray}
 \label{R_function} R&=&\frac{q}{2} \frac{\Gamma(1-\mathrm{i}q)}{\Gamma(1+\mathrm{i}q)}\,\exp\!\left[{2\mathrm{i}q \ln\!\left(\!\sin\frac{\theta}{2}\right)}\right],\\
 \label{D_k_function} D_k&=&-\exp\left(-{\mathrm{i}\pi\xi_k}\right) \, \frac{\Gamma(\xi_k-\mathrm{i}q)}{\Gamma(\xi_k+1+\mathrm{i}q)} +(-)^k\frac{\Gamma(k-\mathrm{i}q)}{\Gamma(k+1+\mathrm{i}q)},\\
\label{D_k_function1}  & =&  -\frac{\exp\left(-{\mathrm{i}\pi\xi_k}\right)}{\xi_k+iq} \, \frac{\Gamma(\xi_k-\mathrm{i}q)}{\Gamma(\xi_k+\mathrm{i}q)} +\frac{\exp\left({-\mathrm{i}\pi k}\right)}{k+iq}\frac{\Gamma(k-\mathrm{i}q)}{\Gamma(k+\mathrm{i}q)},\\
 \label{xi_function} \xi_k&=&\sqrt{k^2-(\alpha')^2},\\
 \label{q_function} q&=& \frac{\alpha'}{\beta} ,\\
 \label{alpha'_function} \alpha'& =& - \alpha zZ,
\end{eqnarray}
$\ln$ being the natural logarithm and $\Gamma(\mu)$ the gamma function with argument $\mu$.
\section{An Improved Numerical Approach for the Unscreened MDCS}
\label{INAUMDCS}
As discussed in Sect.~\ref{MDCR_Sect}, the MDCS [Eq.~(\ref{MDCS_Mott_R})] can be obtained from the evaluation of its ratio [$\mathcal{R}^{\rm Mott}$, see Eqs.~(\ref{R_Mott_Idoeta}--\ref{G_function})] with respect to Rutherford's formula [Eg.~(\ref{Rutherd_c_s_com_plab_n})].
Sherman~(1956) and Idoeta and Legarda~(1992) pointed out that the function $F_1$ and $G_1$ [Eqs.~(\ref{F1_function},~\ref{G1_function})] are obtained from two series which are only conditionally convergent and converge very slowly (in particular $G_1$).~Thus, in order to improve the convergence they suggested to apply to such series the transformation - which can be employed for any series of Legendre polynomials - of Yennie, Ravenhall and Wilson~(1954).~Following this transformation, the series of Legendre polynomials (with the $k$-th term given by $\mathrm{P}_k(\cos\theta)$ in Eqs.~(\ref{F1_function},~\ref{G1_function})] are re-written in terms of an $m$-th reduced series,~i.e.,
\begin{eqnarray}
 \label{F1_function_YRW}F_1&=&\frac{i}{2\left(1-\cos\theta\right)^{m}}\,\sum_{k=0}^{+\infty}A_k^{(m)}\mathrm{P}_k(\cos\theta),\\
 \label{G1_function_YRW}G_1&=&\frac{i}{2\left(1-\cos\theta\right)^{m}}\,\sum_{k=0}^{+\infty}B_k^{(m)}\mathrm{P}_k(\cos\theta),
\end{eqnarray}
with
\begin{eqnarray}
   \label{A_0_YRW}A_k^{(0)}&=&(-)^k[kD_k+(k+1)D_{k+1}],\\
  \nonumber \ldots & = & \ldots,\\
   \nonumber A_k^{(m-1)} &= &A_{k}^{(m-2)}-\left[\frac{k+1}{2k+3}\right]A_{k+1}^{(m-2)}-\left[\frac{k}{2k-1}\right]A_{k-1}^{(m-2)},\\
\label{A_m_YRW}A_k^{(m)}&=&A_{k}^{(m-1)}-\left[\frac{k+1}{2k+3}\right]A_{k+1}^{(m-1)}-\left[\frac{k}{2k-1}\right]A_{k-1}^{(m-1)}
\end{eqnarray}
and
\begin{eqnarray}
 \label{B_0_YRW}B_k^{(0)}&=&(-)^k[k^2D_k-(k+1)^2D_{k+1}],\\
 \nonumber \ldots & = & \ldots,\\
 \nonumber B_k^{(m-1)} &= &B_{k}^{(m-2)}-\left[\frac{k+1}{2k+3}\right]B_{k+1}^{(m-2)}-\left[\frac{k}{2k-1}\right]B_{k-1}^{(m-2)},\\
\label{B_m_YRW} B_k^{(m)}&=&B_{k}^{(m-1)}-\left[\frac{k+1}{2k+3}\right]B_{k+1}^{(m-1)}-\left[\frac{k}{2k-1}\right]B_{k-1}^{(m-1)}
\end{eqnarray}
(e.g.,~see \cite{Yennie,Sherman,Idoeta,Lijian}). Using a code - developed for this purpose - in the Mathematica 8.0 environment~\cite{Mathematica}, three reductions ($m=3$) were found to be adequate for the purpose of present calculations in agreement with~\cite{Sherman,Idoeta}.~Since the series $G_1$ is more slowly convergent than $F_1$, in the same code the \textit{Euler transformation} (e.g., see~Chapter VIII in \cite{Knoff} and \cite{Sherman,Idoeta}) was additionally applied to the so-found 3rd reduced series, i.e.,
\begin{equation}
 \label{Euler_Tr}\sum_{k=0}^{\infty}(-1)^k v_k=\frac{v_0}{2}+\frac{\Delta^1 v_0}{4}+ \dots +\frac{{\Delta^p v_0}}{2^{p+1}}+\sum_{m=0}^{\infty}\frac{(-1)^m \Delta^{p+1} v_m}{2^{p+1}}
\end{equation}
with
\begin{eqnarray}
 \nonumber v_k &=& A_k^{(3)}\mathrm{P}_k(\cos\theta) \left[\frac{i}{(-1)^k \,2\left(1-\cos\theta\right)^{3} }\right],\\
 \nonumber v_m &=& A_m^{(3)}\mathrm{P}_m(\cos\theta) \left[\frac{i}{(-1)^m \,2\left(1-\cos\theta\right)^{3} }\right]
\end{eqnarray}
and
\begin{eqnarray}
 \nonumber \Delta^1 v_m&=&v_m-v_{m+1},\\
  \nonumber \ldots &=& \ldots,\\
 \nonumber \Delta^p v_m &=& \Delta^{p-1} v_m - \Delta^{p-1} v_{m+1},\\
  \nonumber \Delta^{p+1} v_m &=& \Delta^{p} v_m - \Delta^{p} v_{m+1}.
\end{eqnarray}
For the current calculations, $p=1$ was used in Eq.~(\ref{Euler_Tr}).~Finally, since in the developed code no approximation was introduced to evaluate Eqs.~(\ref{F0_function},~\ref{G0_function}), the accuracy in calculating the $\mathcal{R}^{\rm Mott}$ ratio [as function of $p$, $\theta$, $Z$ and $z$,~e.g., see Eq.~(\ref{R_Mott_Idoeta})] depends on the accuracy with which Eqs.~(\ref{F_function},~\ref{G_function}) are calculated.~Thus, in turn, it results from the number (i.e.,~the value of $k$) of additive terms - approximated as so far discussed - summed in Eqs.~(\ref{F1_function},~\ref{G1_function}).~The result was considered \textit{accurate} when the last $k$-th term summed was such that the obtained $\mathcal{R}^{\rm Mott}$ value [$\equiv R^{\rm num} (\theta, Z, E)]$ varies by less than $10^{-6}$ when seven more terms in the series -~i.e., extending the sums up to $k+7$ - were added.
\begin{table}
\tiny
\caption{MDCS i) calculated following the approach discussed in Sect.~\ref{INAUMDCS} and ii) from Idoeta and Legarda~(1992) - indicated as I\&L~(1992) - for electrons at 5\,keV, 500\,keV and 10\,MeV on He, Ag and U ($Z= 2, \, 47 \, \textrm{and} \, 92$, respectively) as a function of the scattering angle $\theta$ (in deg).}
\centering%
\begin{tabular}{r|r|rr|rr|rr}
\toprule%
  &       & \multicolumn{2}{c|}{5\,keV} & \multicolumn{2}{c|}{500\,keV} & \multicolumn{2}{c}{10\,MeV} \\
$Z$ & $\theta$ (in deg) & Present work & I\&L~(1992) & Present work & I\&L~(1992)& Present work & I\&L~(1992) \\
\toprule%
	&	15	&	1.00044E+00	&	1.00044E+00	&	9.91871E-01	&	9.91873E-01	&	9.88268E-01	&	9.88269E-01	\\
	&	30	&	1.00006E+00	&	1.00006E+00	&	9.57871E-01	&	9.57871E-01	&	9.42117E-01	&	9.42117E-01	\\
	&	45	&	9.98883E-01	&	9.98883E-01	&	9.00541E-01	&	9.00541E-01	&	8.64954E-01	&	8.64953E-01	\\
	&	60	&	9.97029E-01	&	9.97029E-01	&	8.24032E-01	&	8.24032E-01	&	7.62318E-01	&	7.62318E-01	\\
	&	75	&	9.94669E-01	&	9.94669E-01	&	7.33796E-01	&	7.33796E-01	&	6.41480E-01	&	6.41479E-01	\\
2	&	90	&	9.92004E-01	&	9.92004E-01	&	6.36207E-01	&	6.36207E-01	&	5.10931E-01	&	5.10931E-01	\\
	&	105	&	9.89254E-01	&	9.89254E-01	&	5.38120E-01	&	5.38120E-01	&	3.79805E-01	&	3.79805E-01	\\
	&	120	&	9.86639E-01	&	9.86638E-01	&	4.46399E-01	&	4.46399E-01	&	2.57246E-01	&	2.57246E-01	\\
	&	135	&	9.84362E-01	&	9.84362E-01	&	3.67449E-01	&	3.67449E-01	&	1.51785E-01	&	1.51784E-01	\\
	&	150	&	9.82599E-01	&	9.82599E-01	&	3.06771E-01	&	3.06772E-01	&	7.07477E-02	&	7.07482E-02	\\
	&	165	&	9.81485E-01	&	9.81485E-01	&	2.68590E-01	&	2.68590E-01	&	1.97618E-02	&	1.97617E-02	\\
	&	180	&	9.81104E-01	&	9.81104E-01	&	2.55561E-01	&	2.55561E-01	&	2.36433E-03	&	2.36435E-03	\\\midrule
	&	15	&	1.00121E+00	&	1.00121E+00	&	1.09603E+00	&	1.09602E+00	&	1.11558E+00	&	1.11557E+00	\\
	&	30	&	1.00638E+00	&	1.00638E+00	&	1.18990E+00	&	1.18989E+00	&	1.21467E+00	&	1.21466E+00	\\
	&	45	&	9.96364E-01	&	9.96365E-01	&	1.24149E+00	&	1.24149E+00	&	1.25631E+00	&	1.25631E+00	\\
	&	60	&	9.90869E-01	&	9.90869E-01	&	1.23525E+00	&	1.23525E+00	&	1.22547E+00	&	1.22547E+00	\\
	&	75	&	1.00217E+00	&	1.00217E+00	&	1.16997E+00	&	1.16997E+00	&	1.12221E+00	&	1.12221E+00	\\
47	&	90	&	1.03381E+00	&	1.03381E+00	&	1.05458E+00	&	1.05458E+00	&	9.58058E-01	&	9.58058E-01	\\
	&	105	&	1.08620E+00	&	1.08620E+00	&	9.05084E-01	&	9.05084E-01	&	7.53051E-01	&	7.53051E-01	\\
	&	120	&	1.15388E+00	&	1.15388E+00	&	7.41682E-01	&	7.41682E-01	&	5.32623E-01	&	5.32623E-01	\\
	&	135	&	1.22582E+00	&	1.22582E+00	&	5.85909E-01	&	5.85909E-01	&	3.24172E-01	&	3.24172E-01	\\
	&	150	&	1.28896E+00	&	1.28896E+00	&	4.57908E-01	&	4.57909E-01	&	1.53572E-01	&	1.53572E-01	\\
	&	165	&	1.33186E+00	&	1.33186E+00	&	3.74006E-01	&	3.74006E-01	&	4.19648E-02	&	4.19648E-02	\\
	&	180	&	1.34702E+00	&	1.34702E+00	&	3.44813E-01	&	3.44813E-01	&	3.16260E-03	&	3.16261E-03	\\\midrule
	&	15	&	1.00302E+00	&	1.00310E+00	&	1.05841E+00	&	1.05837E+00	&	1.09636E+00	&	1.09636E+00	\\
	&	30	&	1.00860E+00	&	1.00859E+00	&	1.20318E+00	&	1.20319E+00	&	1.32107E+00	&	1.32107E+00	\\
	&	45	&	9.90926E-01	&	9.90927E-01	&	1.50713E+00	&	1.50713E+00	&	1.70264E+00	&	1.70264E+00	\\
	&	60	&	1.02493E+00	&	1.02492E+00	&	1.87820E+00	&	1.87820E+00	&	2.11844E+00	&	2.11844E+00	\\
	&	75	&	1.05048E+00	&	1.05048E+00	&	2.18903E+00	&	2.18903E+00	&	2.41683E+00	&	2.41683E+00	\\
92	&	90	&	9.25197E-01	&	9.25197E-01	&	2.33938E+00	&	2.33938E+00	&	2.47982E+00	&	2.47982E+00	\\
	&	105	&	8.08367E-01	&	8.08368E-01	&	2.28576E+00	&	2.28576E+00	&	2.26008E+00	&	2.26008E+00	\\
	&	120	&	9.44562E-01	&	9.44562E-01	&	2.04832E+00	&	2.04832E+00	&	1.79365E+00	&	1.79365E+00	\\
	&	135	&	1.36440E+00	&	1.36440E+00	&	1.69988E+00	&	1.69988E+00	&	1.18892E+00	&	1.18892E+00	\\
	&	150	&	1.89411E+00	&	1.89411E+00	&	1.34241E+00	&	1.34241E+00	&	5.96098E-01	&	5.96098E-01	\\
	&	165	&	2.31435E+00	&	2.31435E+00	&	1.07789E+00	&	1.07789E+00	&	1.65995E-01	&	1.65995E-01	\\
	&	180	&	2.47185E+00	&	2.47185E+00	&	9.80670E-01	&	9.80671E-01	&	9.13938E-03	&	9.13939E-03	\\\bottomrule
\end{tabular}
\label{E1}
\normalsize
\end{table}
\par
Idoeta and Legarda~(1992) reported their calculated MDCS for electrons and positrons from 5\,keV up to 10\,MeV on a few nuclei with $Z$ from 2 up to 92 as function of the scattering angle ($\theta$, every $15^{\circ}$ from $15^{\circ}$ up $180^{\circ}$).~In addition, they already discussed the agreement of their results with those from i) the Mott--Born formula obtained in the first-order Born approximation (e.g., see \cite{Motz_rmp}), ii) McKinley--Feshbach expression derived using the second-order Born approximation \cite{McKinley,Motz_rmp}), iii) Curr formula derived in the $(\alpha Z)^5$ and (for $\beta \simeq 1$)  $(\alpha Z)^8$ approximations (e.g., see \cite{Curr}) and, finally, iv) from Refs.~\cite{Yadav,Doddett,Sherman}.
\par
\begin{figure}
\begin{center}
 \includegraphics[width=.85\textwidth]{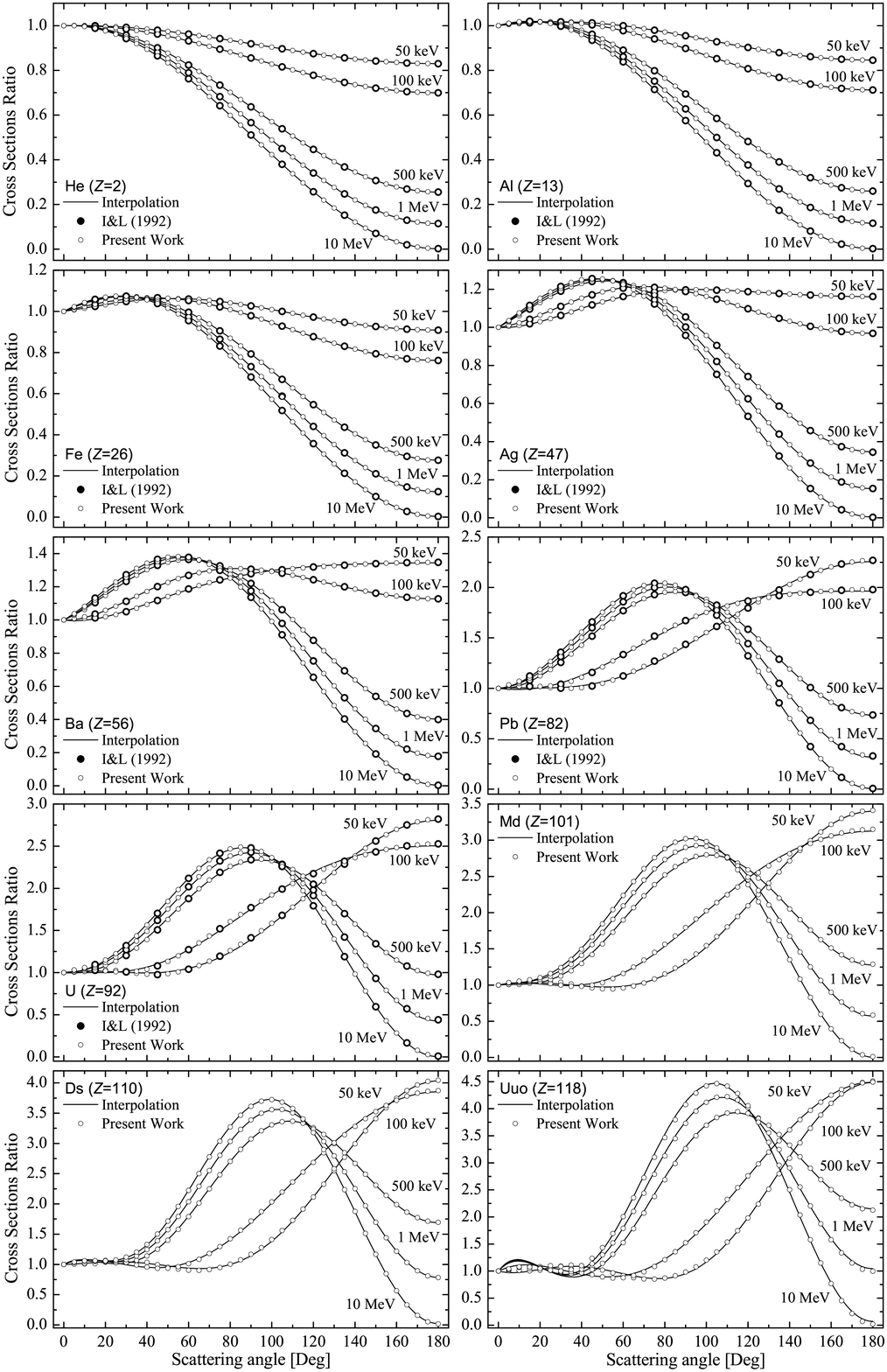}
 \vskip -.4cm
 \caption{Cross section ratio (indicated with $\mathcal{R}^{\rm Mott}$ in the text) as function of scattering angle (in degrees) for electrons with 50\,keV, 100\,keV, 500\,keV, 1\,MeV and 10\,MeV interacting on He, Al, Fe, Ag, Ba, Pb, U, Md, Ds and Uuo: the interpolated curves (continous lines) were obtained from Eq.~(\ref{Mott_R_interp_expr}), ($\circ$) from the improved numerical approach discussed in Sect.~\ref{INAUMDCS} and, finally, ($\bullet$) from Idoeta and Legarda~(1992) - indicated as I\&L~(1992) -.}\label{comp_Electrons}
\end{center}
\end{figure}
\begin{figure}
\begin{center}
 \includegraphics[width=.85\textwidth]{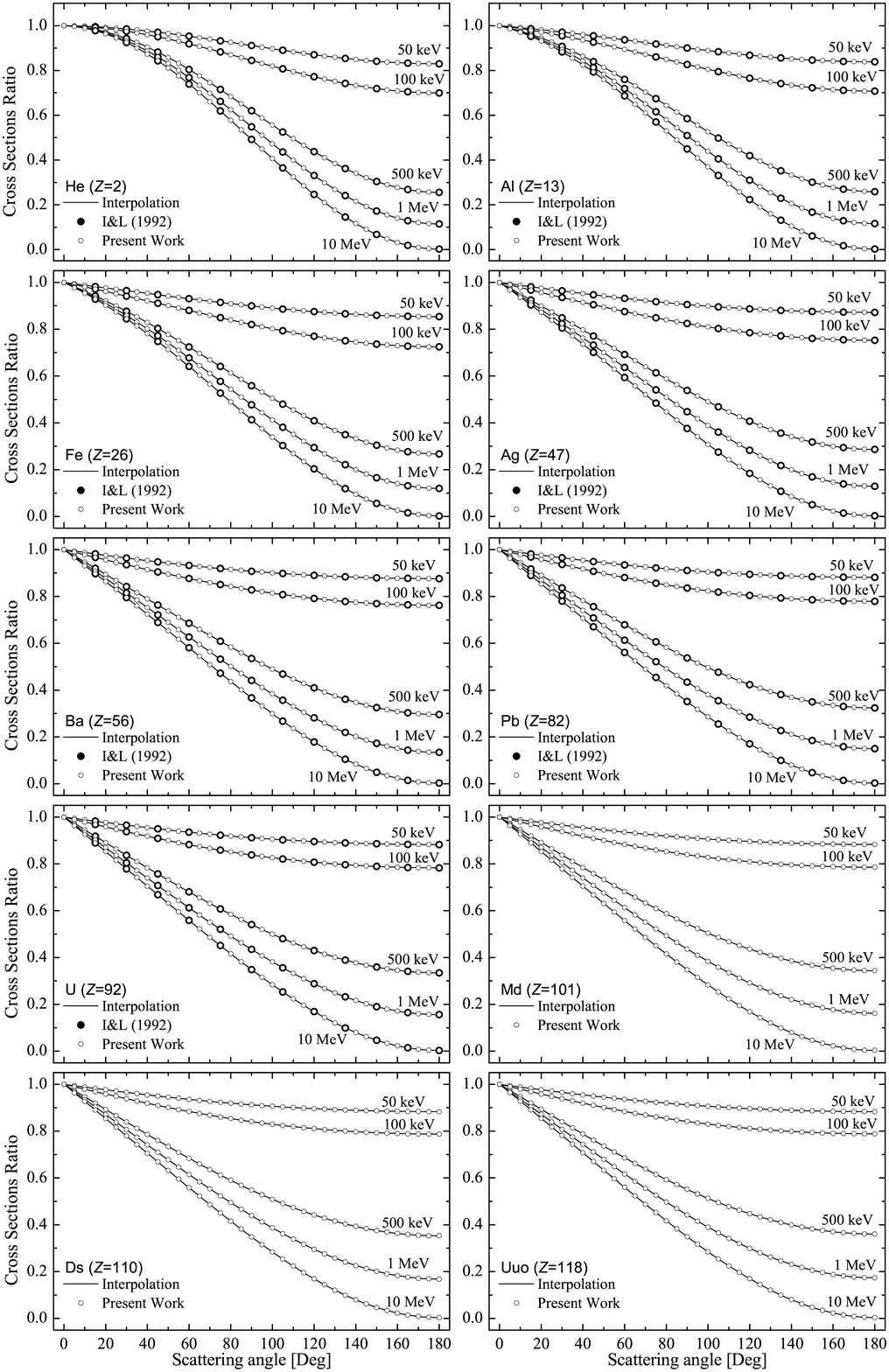}
  \vskip -.4cm
\caption{Cross section ratio as function of scattering angle (in degrees) for positrons with 50\,keV, 100\,keV, 500\,keV, 1\,MeV and 10\,MeV interacting on He, Al, Fe, Ag, Ba, Pb, U, Md, Ds and Uuo: the interpolated curves (continous lines) were obtained from Eq.~(\ref{Mott_R_interp_expr}), ($\circ$) from the improved numerical approach discussed in Sect.~\ref{INAUMDCS} and, finally, ($\bullet$) from Idoeta and Legarda~(1992) - indicated as I\&L~(1992) -.}\label{comp_Positrons}
\end{center}
\end{figure}

The presently obtained values were compared with those determined by Idoeta and Legarda~(1992) and found to agree at least up to the 3rd digit (occurring in the 0.1\% of the cases) or more,~e.g., up to the 6th digit in the 73\% of the cases (see also Figs.~\ref{comp_Electrons} and~\ref{comp_Positrons}).~For instance, in Table~\ref{E1} (Table~\ref{P1}) the MDCS for electrons (positrons) on three nuclei - He, Ag and U ($Z= 2, \, 47 \, \textrm{and} \, 92$, respectively) - are shown as function of the the scattering angle $\theta$ (in deg) at 5\,keV (the lowest energy treated in Ref.~\cite{Idoeta}), 500\,keV and 10\,MeV (the largest energy treated in Ref.~\cite{Idoeta}).
\begin{table}
\tiny
\caption{MDCS i) calculated following the approach discussed in Sect.~\ref{INAUMDCS} and ii) from Idoeta and Legarda~(1992) - indicated as I\&L~(1992) - for positrons at 5\,keV, 500\,keV and 10\,MeV on He, Ag and U ($Z= 2, \, 47 \, \textrm{and} \, 92$, respectively) as a function of the scattering angle $\theta$ (in deg).}
\centering%
\begin{tabular}{r|r|rr|rr|rr}
\toprule%
  &       & \multicolumn{2}{c|}{5\,keV} & \multicolumn{2}{c|}{500\,keV} & \multicolumn{2}{c}{10\,MeV} \\
$Z$ & $\theta$ (in deg) & Present work & I\&L~(1992) & Present work & I\&L~(1992)& Present work & I\&L~(1992) \\
\toprule%
	&	15	&	9.99020E-01	&	9.99020E-01	&	9.82895E-01	&	9.82895E-01	&	9.77877E-01	&	9.77874E-01	\\
	&	30	&	9.97635E-01	&	9.97634E-01	&	9.42692E-01	&	9.42693E-01	&	9.24546E-01	&	9.24546E-01	\\
	&	45	&	9.95880E-01	&	9.95880E-01	&	8.81845E-01	&	8.81845E-01	&	8.43312E-01	&	8.43312E-01	\\
	&	60	&	9.93839E-01	&	9.93839E-01	&	8.04244E-01	&	8.04244E-01	&	7.39413E-01	&	7.39412E-01	\\
	&	75	&	9.91616E-01	&	9.91616E-01	&	7.14942E-01	&	7.14942E-01	&	6.19656E-01	&	6.19655E-01	\\
2	&	90	&	9.89333E-01	&	9.89333E-01	&	6.19810E-01	&	6.19810E-01	&	4.91952E-01	&	4.91952E-01	\\
	&	105	&	9.87119E-01	&	9.87119E-01	&	5.25138E-01	&	5.25139E-01	&	3.64780E-01	&	3.64781E-01	\\
	&	120	&	9.85100E-01	&	9.85099E-01	&	4.37211E-01	&	4.37210E-01	&	2.46613E-01	&	2.46612E-01	\\
	&	135	&	9.83392E-01	&	9.83392E-01	&	3.61878E-01	&	3.61878E-01	&	1.45339E-01	&	1.45339E-01	\\
	&	150	&	9.82096E-01	&	9.82096E-01	&	3.04162E-01	&	3.04163E-01	&	6.77310E-02	&	6.77317E-02	\\
	&	165	&	9.81286E-01	&	9.81286E-01	&	2.67914E-01	&	2.67915E-01	&	1.89844E-02	&	1.89843E-02	\\
	&	180	&	9.81011E-01	&	9.81011E-01	&	2.55557E-01	&	2.55557E-01	&	2.36430E-03	&	2.36432E-03	\\\midrule
	&	15	&	9.98027E-01	&	9.98026E-01	&	9.24324E-01	&	9.24325E-01	&	9.03683E-01	&	9.03684E-01	\\
	&	30	&	9.96141E-01	&	9.96142E-01	&	8.48536E-01	&	8.48536E-01	&	8.05107E-01	&	8.05107E-01	\\
	&	45	&	9.94405E-01	&	9.94405E-01	&	7.70920E-01	&	7.70920E-01	&	7.01517E-01	&	7.01517E-01	\\
	&	60	&	9.92842E-01	&	9.92842E-01	&	6.92096E-01	&	6.92096E-01	&	5.93677E-01	&	5.93677E-01	\\
	&	75	&	9.91466E-01	&	9.91466E-01	&	6.13800E-01	&	6.13800E-01	&	4.84102E-01	&	4.84101E-01	\\
47	&	90	&	9.90279E-01	&	9.90280E-01	&	5.38380E-01	&	5.38380E-01	&	3.76398E-01	&	3.76398E-01	\\
	&	105	&	9.89283E-01	&	9.89283E-01	&	4.68481E-01	&	4.68481E-01	&	2.74816E-01	&	2.74816E-01	\\
	&	120	&	9.88474E-01	&	9.88475E-01	&	4.06791E-01	&	4.06791E-01	&	1.83837E-01	&	1.83837E-01	\\
	&	135	&	9.87849E-01	&	9.87850E-01	&	3.55825E-01	&	3.55825E-01	&	1.07779E-01	&	1.07779E-01	\\
	&	150	&	9.87405E-01	&	9.87405E-01	&	3.17743E-01	&	3.17743E-01	&	5.04405E-02	&	5.04406E-02	\\
	&	165	&	9.87140E-01	&	9.87140E-01	&	2.94200E-01	&	2.94200E-01	&	1.47809E-02	&	1.47809E-02	\\
	&	180	&	9.87052E-01	&	9.87051E-01	&	2.86235E-01	&	2.86235E-01	&	2.67975E-03	&	2.67977E-03	\\\midrule
	&	15	&	9.98030E-01	&	9.97955E-01	&	9.18244E-01	&	9.18243E-01	&	8.89429E-01	&	8.89428E-01	\\
	&	30	&	9.96160E-01	&	9.96158E-01	&	8.36871E-01	&	8.63871E-01	&	7.78709E-01	&	7.78709E-01	\\
	&	45	&	9.94432E-01	&	9.94432E-01	&	7.57144E-01	&	7.57144E-01	&	6.68219E-01	&	6.68218E-01	\\
	&	60	&	9.92877E-01	&	9.92877E-01	&	6.80232E-01	&	6.80232E-01	&	5.58530E-01	&	5.58530E-01	\\
	&	75	&	9.91507E-01	&	9.91507E-01	&	6.07527E-01	&	6.07527E-01	&	4.51125E-01	&	4.51125E-01	\\
92	&	90	&	9.90325E-01	&	9.90325E-01	&	5.40566E-01	&	5.40466E-01	&	3.48350E-01	&	3.48349E-01	\\
	&	105	&	9.89331E-01	&	9.89332E-01	&	4.80894E-01	&	4.80894E-01	&	2.53196E-01	&	2.53196E-01	\\
	&	120	&	9.88524E-01	&	9.88524E-01	&	4.29951E-01	&	4.29951E-01	&	1.69019E-01	&	1.69018E-01	\\
	&	135	&	9.87899E-01	&	9.87899E-01	&	3.88984E-01	&	3.88984E-01	&	9.92006E-02	&	9.92008E-02	\\
	&	150	&	9.87455E-01	&	9.87456E-01	&	3.58994E-01	&	3.59884E-01	&	4.68185E-02	&	4.68182E-02	\\
	&	165	&	9.87190E-01	&	9.87190E-01	&	3.40708E-01	&	3.40708E-01	&	1.43292E-02	&	1.43291E-02	\\
	&	180	&	9.87102E-01	&	9.87102E-01	&	3.34563E-01	&	3.34564E-01	&	3.31703E-03	&	3.31705E-03	\\\bottomrule
\end{tabular}
\label{P1}
\normalsize
\end{table}
%
\section{A Practical Interpolated Expression for the Unscreened MDCS}
\label{PIEUMDCS}
Recently, Lijian, Quing and Zhengming~(1995) suggested a practical interpolated polynomial expression [Eq.~(\ref{Mott_R_interp_expr})] to $ \mathcal{R}^{\rm Mott}$ [Eq.~(\ref{R_Mott_Idoeta})].~The expression was a function of both $\theta$ and $\beta$ for electrons with energies from 1\,keV up to 900\,MeV interacting on nuclei with $1 \leqslant Z \leqslant 90$,~i.e.,
\begin{equation}\label{Mott_R_interp_expr}
  \mathcal{R}^{\rm Mott}\equiv R^{\rm int}(\theta, Z, E) = \sum_{\textrm{j}=0}^4 a_\textrm{j}(Z,\beta)(1-\cos\theta)^{\textrm{j}/2},
\end{equation}
where
\begin{equation}\label{Mott_R_interp_expr_coeff}
a_\textrm{j}(Z,\beta)=\sum_{\textrm{k}=1}^6 b_\textrm{k,j}(Z)(\beta-\overline{\beta})^{\textrm{k}-1},
\end{equation}
and
$\overline{\beta}\,c=0.7181287 \,c$ is the mean velocity of electrons within the above mentioned energy range.~The coefficients $b_\textrm{k,j}(Z)$ obtained by Lijian, Quing and Zhengming~(1995) are listed in~Table 1 of their article.~Furthermore, it has to be pointed out that the energy dependence of $\mathcal{R}^{\rm Mott}$ from Eq.~(\ref{Mott_R_interp_expr}) was studied and observed to be negligible above $\approx 10\,$MeV (as expected from Eq.~(\ref{Mott_R_interp_expr_coeff}), because $\beta$ approaches 1, and discussed in \cite{Lijian}).~It has to be remarked that at 10, 100 and 1000\,MeV for Li, Si, Fe and Pb, Boschini and collaborators~(2012) calculated values of $\mathcal{R}^{\rm Mott}$ using both Curr~(1955) and Lijian, Quing and Zhengming~(1995) methods and found them to be in a very good agreement.
\begin{figure}[h]
\begin{center}
 \includegraphics[width=.8\textwidth]{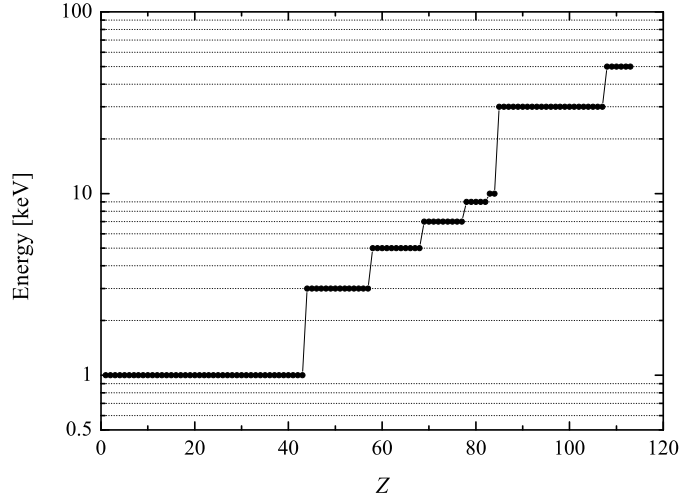}
  \vskip -.4cm
\caption{Energy (in keV) of electrons above which the interpolated expression~(\ref{Mott_R_interp_expr}) provides values of $\mathcal{R}^{\rm Mott}$ ratio with $\left< \sigma_{\rm rel} \right> \leq1\%$ as a function of the atomic number $Z$ of target nucleus with $1\leqslant Z \leqslant 114$.}\label{Energy_more_1percent}
\end{center}
\end{figure}
%
\begin{figure}
\begin{center}
 \includegraphics[width=.9\textwidth]{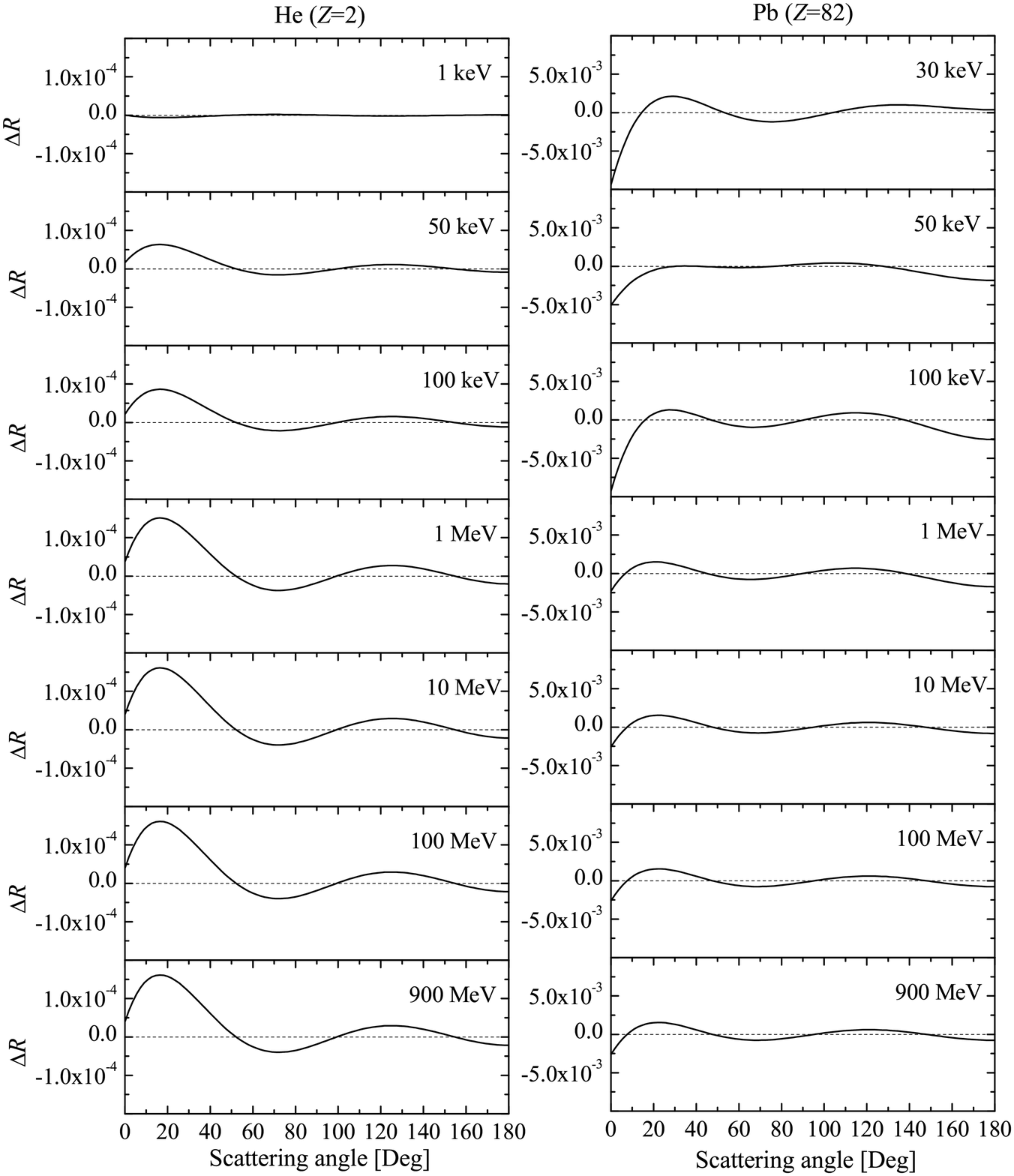}
 \vskip -.4cm
 \caption{Difference, $ \Delta R$, between the interpolated expression~(\ref{Mott_R_interp_expr}) obtained in the present work from that by Lijian, Quing and Zhengming~(1995) as a function of the scattering angle, in degrees, for electrons at 1\,keV (30\,keV), 50\,keV, 100\,keV, 1\,MeV, 10\,MeV, 100\,MeV and 900\,MeV on He (Pb) nuclei.}\label{Diff_from_Lijian}
\end{center}
\end{figure}
\par
In the current work, following the same procedure indicated by Lijian, Quing and Zhengming~(1995), the coefficients $b_\textrm{k,j}(Z)$ for Eqs.(\ref{Mott_R_interp_expr},~\ref{Mott_R_interp_expr_coeff}) were obtained for both electrons (Table~\ref{table_electrons}) and positrons (Table~\ref{table_positrons}) interacting on nuclei with $1 \leqslant Z \leqslant 118$.~Thus extending the treatment to positrons (not discussed in \cite{Lijian}) and to the interaction of electrons with higher $Z$ nuclei. It is worthwhile to remark that the previously available coefficients for electrons, as already noted by Jun and collaborators~(2009), exhibit many typographic errors among those listed in Table 1 of \cite{Lijian}.
\par
For instance, in Fig.~\ref{comp_Electrons} (Fig.~\ref{comp_Positrons}) the interpolated curves, obtained from Eq.~(\ref{Mott_R_interp_expr}), are shown as a function of the scattering angle for electrons (positrons) with 50\,keV, 100\,keV, 500\,keV, 1\,MeV and 10\,MeV interacting on He, Al, Fe, Ag, Ba, Pb, U, Md, Ds and Uuo.~In Figs.~\ref{comp_Electrons} and~\ref{comp_Positrons}, the values from a) the improved numerical approach calculated in the present work (see discussion in Sect.~\ref{INAUMDCS}) and b), when available, from Idoeta and Legarda~(1992) are also shown and found in good agreement with interpolated data.
\par
The precision of the current procedure for interpolation can be treated following that dealt by Lijian, Quing and Zhengming~(1995) as a function of the energy, $E$, of incoming electrons or positrons and the atomic number $Z$ of target nuclei, i.e., introducing the \textit{average relative error}, $\left< \sigma_{\rm rel} \right>$, given by:
\begin{equation}\label{ARE}
\left< \sigma_{\rm rel} \right>= \sqrt{\frac{\sum _{i= 0}^{36}    \left[R^{\rm int}(\theta_i, Z, E) - R^{\rm num} (\theta_i, Z, E)\right]^2}{\sum _{j= 0}^{36}R^{\rm num} (\theta_j, Z, E)^2 } }
\end{equation}
where $\theta_i$ ($\theta_j$) is the $i$th ($j$th) value of the scattering angle from $0^{\circ}$ up to $180^{\circ}$ with $5^{\circ}$ step, $R^{\rm int}$ is the value obtained from Eq.~(\ref{Mott_R_interp_expr}) and, finally,
$R^{\rm num}$ is that from the numerical solution - sometime referred to as \textit{exact solution} [e.g.,~see  Idoeta and Legarda~(1992), and Lijian, Quing and Zhengming~(1995)] - discussed in Sect.~\ref{INAUMDCS}.~For electrons, one finds that, for $1\leqslant Z \leqslant 114$, the interpolated expression~(\ref{Mott_R_interp_expr}) can be used with $\left< \sigma_{\rm rel} \right> \leq1\%$ at energies (shown in Fig.~\ref{Energy_more_1percent}) which increase with increasing $Z$, i.e., the atomic number of the target nucleus; for $Z >114$, $\left< \sigma_{\rm rel} \right>$ is (1--2.3)\% for energies larger than 50\,keV.~For positrons, $\left< \sigma_{\rm rel} \right>$ was found to be less than 0.05\% for all energies (i.e., from 1\,keV up to 900\,MeV) and all nuclei (i.e., $1\leqslant Z \leqslant 118$).
\par
For electrons, the interpolated expression~(\ref{Mott_R_interp_expr}) [$R^{\rm int}(\theta, Z, E)$] obtained in the present work was also compared - when available and usable - with that from Lijian, Quing and Zhengming~(1995) [$ R^{\rm L,Q,Z~(1995)} (\theta, Z, E)$].~The two expressions were found in good agreement with typical differences,
\begin{equation}\label{deltaR}
 \Delta R =  R^{\rm int}(\theta, Z, E) - R^{\rm L,Q,Z~(1995)} (\theta, Z, E),
\end{equation}
not exceeding $10^{-3}$ for low-$Z$ nuclei and $10^{-2}$ for high-$Z$ nuclei.~As an example, Fig.~\ref{Diff_from_Lijian} shows the values of $ \Delta R$ [Eq.~(\ref{deltaR})] obtained for the scattering of electrons on He and Pb nuclei as a function of the scattering angle (in degrees) at the minimum energy for which $\left< \sigma_{\rm rel} \right> \leq 1\%$ -~i.e., 1\,keV for He and 30\,keV for Pb - and at 50\,keV, 100\,keV, 1\,MeV, 10\,MeV, 100\,MeV and 900\,MeV.
%
\newpage
\tiny
\begin{center}
\begin{longtable}{lcrrrrrr}
\caption[Table1]{$b_{\textrm{j,k}}$ with $\textrm{j}=0,4$ and $\textrm{k}=1,6$ coefficients for electrons} \label{table_electrons} \\

\hline \multicolumn{1}{l}{} & \multicolumn{1}{c}{{$b_{\textrm{j,k}}$}} & \multicolumn{1}{c}{} & \multicolumn{1}{c}{} & \multicolumn{1}{c}{} & \multicolumn{1}{c}{} & \multicolumn{1}{c}{} & \multicolumn{1}{c}{} \\
\multicolumn{1}{l}{} & \multicolumn{1}{c}{{j}} & \multicolumn{1}{c}{{k=1}} & \multicolumn{1}{c}{{k=2}} & \multicolumn{1}{c}{{k=3}} & \multicolumn{1}{c}{{k=4}} & \multicolumn{1}{c}{{k=5}} & \multicolumn{1}{c}{{k=6}} \\ \hline \\
\endfirsthead

\multicolumn{7}{c}%
{{\tablename\ \thetable{} -- continued from previous page}} \\
\hline \multicolumn{1}{l}{} & \multicolumn{1}{c}{{$b_{\textrm{j,k}}$}} & \multicolumn{1}{c}{} & \multicolumn{1}{c}{} & \multicolumn{1}{c}{} & \multicolumn{1}{c}{} & \multicolumn{1}{c}{} & \multicolumn{1}{c}{} \\
\multicolumn{1}{l}{} & \multicolumn{1}{c}{{j}} & \multicolumn{1}{c}{{k=1}} & \multicolumn{1}{c}{{k=2}} & \multicolumn{1}{c}{{k=3}} & \multicolumn{1}{c}{{k=4}} & \multicolumn{1}{c}{{k=5}} & \multicolumn{1}{c}{{k=6}} \\ \hline
\endhead

\multicolumn{3}{l}{{Continued on next page}} \\
\hline
\endfoot

\hline
\endlastfoot
& 0 & 1.00000E+00 & 2.67363E-08 & 7.11530E-08 & -9.77030E-08 & -6.69132E-07 & -3.09263E-07 \\
 & 1 & 1.17182E-02 & 1.62222E-02 & -5.90397E-05 & -1.05585E-04 & 4.17873E-04 & 9.13843E-04 \\
H & 2 & -2.65955E-01 & -7.29531E-01 & -4.99796E-01 & 2.83507E-04 & -9.09042E-04 & -2.20244E-03 \\
 & 3 & -1.82348E-04 & -8.86355E-05 & -1.90554E-04 & -2.49708E-04 & 6.35004E-04 & 1.73523E-03 \\
 & 4 & 4.70966E-05 & -4.09705E-06 & 3.75218E-05 & 8.05645E-05 & -1.90534E-04 & -5.42847E-04 \\
 &  & \multicolumn{1}{l}{} & \multicolumn{1}{l}{} & \multicolumn{1}{l}{} & \multicolumn{1}{l}{} & \multicolumn{1}{l}{} & \multicolumn{1}{l}{} \\
 & 0 & 1.00000E+00 & 3.76476E-08 & -3.05313E-07 & -3.27422E-07 & 2.44235E-06 & 4.08754E-06 \\
 & 1 & 2.35767E-02 & 3.24642E-02 & -6.37269E-04 & -7.69160E-04 & 5.28004E-03 & 9.45642E-03 \\
\hline
\end{longtable}
\end{center}
\normalsize

\newpage

\tiny
\begin{center}
\begin{longtable}{lcrrrrrr}
\caption[Table1]{$b_{\textrm{j,k}}$ with $\textrm{j}=0,4$ and $\textrm{k}=1,6$ coefficients for positrons} \label{table_positrons} \\

\hline \multicolumn{1}{l}{} & \multicolumn{1}{c}{{$b_{\textrm{j,k}}$}} & \multicolumn{1}{c}{} & \multicolumn{1}{c}{} & \multicolumn{1}{c}{} & \multicolumn{1}{c}{} & \multicolumn{1}{c}{} & \multicolumn{1}{c}{} \\
\multicolumn{1}{l}{} & \multicolumn{1}{c}{{j}} & \multicolumn{1}{c}{{k=1}} & \multicolumn{1}{c}{{k=2}} & \multicolumn{1}{c}{{k=3}} & \multicolumn{1}{c}{{k=4}} & \multicolumn{1}{c}{{k=5}} & \multicolumn{1}{c}{{k=6}} \\ \hline \\
\endfirsthead

\multicolumn{7}{c}%
{{\tablename\ \thetable{} -- continued from previous page}} \\
\hline \multicolumn{1}{l}{} & \multicolumn{1}{c}{{$b_{\textrm{j,k}}$}} & \multicolumn{1}{c}{} & \multicolumn{1}{c}{} & \multicolumn{1}{c}{} & \multicolumn{1}{c}{} & \multicolumn{1}{c}{} & \multicolumn{1}{c}{} \\
\multicolumn{1}{l}{} & \multicolumn{1}{c}{{j}} & \multicolumn{1}{c}{{k=1}} & \multicolumn{1}{c}{{k=2}} & \multicolumn{1}{c}{{k=3}} & \multicolumn{1}{c}{{k=4}} & \multicolumn{1}{c}{{k=5}} & \multicolumn{1}{c}{{k=6}} \\ \hline
\endhead

\multicolumn{3}{l}{{Continued on next page}} \\
\hline
\endfoot

\hline
\endlastfoot
 & 0 & 1.00000E+00 & -2.63961E-08 & -1.05983E-07 & 6.98859E-08 & 9.77498E-07 & 7.22885E-07 \\
 & 1 & -1.15697E-02 & -1.62210E-02 & 2.56754E-05 & 5.94651E-05 & -1.25966E-04 & -4.19100E-04 \\
H & 2 & -2.49475E-01 & -7.06540E-01 & -5.00006E-01 & -1.93135E-04 & 3.28434E-04 & 1.24240E-03 \\
 & 3 & -2.01733E-04 & -1.52457E-04 & -3.45949E-05 & 1.75794E-04 & -1.53568E-04 & -9.52819E-04 \\
 & 4 & 5.43511E-05 & 1.80300E-05 & -9.61947E-06 & -5.72438E-05 & 3.75348E-05 & 2.95392E-04 \\
 &  & \multicolumn{1}{l}{} & \multicolumn{1}{l}{} & \multicolumn{1}{l}{} & \multicolumn{1}{l}{} & \multicolumn{1}{l}{} & \multicolumn{1}{l}{} \\
 & 0 & 1.00000E+00 & -3.71563E-08 & 2.54023E-07 & 2.73203E-07 & -1.93946E-06 & -3.30422E-06 \\
 & 1 & -2.29788E-02 & -3.24498E-02 & 2.27671E-04 & 1.90496E-04 & -1.58887E-03 & -3.00192E-03 \\
\hline
\end{longtable}
\end{center}
\normalsize
%
\section{Conclusion}
In the present work, an improved numerical approach for determining the
ratio,$\mathcal{R}^{\rm Mott}$, of the unscreened MDCS with respect to Rutherford's formula
was discussed, compared with that from Idoeta and Legarda~(1992) up to 10\,MeV and, finally, found to agree at least up to the 3rd digit (occurring in the 0.1\% of the cases) or more,~e.g., up to the 6th digit in the 73\% of the cases.
\par
For both electrons and positrons scattering on nuclei with $1\leqslant Z \leqslant 118$ in the energy range from 1\,keV up to 900\,MeV, the calculated numerical values of $\mathcal{R}^{\rm Mott}$ from the current improved numerical solution were used to provide an interpolated practical expression.~The latter function was shown to be in good agreement with the numerical solution.~In addition, this interpolated expression was also compared with that found previously - when available and usable - only for electrons (on nuclei with $1\leqslant Z \leqslant 90$) by Lijian, Quing and Zhengming~(1995).~The two expressions exhibit differences not exceeding $10^{-3}$ for low-$Z$ nuclei and $10^{-2}$ for high-$Z$ nuclei.

\section*{Acknowledges}

The work is supported by ASI (Italian Space Agency) under contract ASI-INFN 075/09/0 and ESA (European Space Agency) under ESTEC contract 4000103268.

\bibliographystyle{elsarticle-harv}







\end{document}